\documentclass{article}
\usepackage{spconf,amsmath,amssymb,graphicx,booktabs,hyperref,xcolor}
\hypersetup{hidelinks}

\newcommand{\model}{Nemotron~3 Diarization}

\title{Nemotron 3 Diarization: Advancing the Accuracy--Efficiency Frontier \linebreak in Streaming Speaker Diarization}

\name{\begin{tabular}{@{}c@{}}
  Ivan Medennikov$^{1}$, Taejin Park$^{1}$, Tatiana Timofeeva$^{1}$, He Huang$^{1}$, Alexander Polok$^{1,2}$, \\
  Sofia Kostandian$^{1}$, Weiqing Wang$^{1}$, Dongji Gao$^{1}$, Jagadeesh Balam$^{1}$, Boris Ginsburg$^{1}$
  \end{tabular}
}

\address{
  $^{1}$NVIDIA, USA \quad 
  $^{2}$Brno University of Technology, Czech Republic \quad
}

\makeatletter
\let\originalthebibliography\thebibliography
\renewcommand{\thebibliography}[1]{%
  \originalthebibliography{#1}%
  \setlength{\itemsep}{0pt}%
}
\makeatother

\begin{document}
\ninept
\maketitle

\begin{abstract}
Streaming Sortformer introduced an arrival-order speaker cache to maintain speaker identities across chunks in end-to-end streaming speaker diarization, with a single model supporting configurable latency profiles spanning offline operation to ultra-low-latency streaming.
Nemotron 3 Diarization addresses the original model’s four-speaker limitation by extending support to eight speakers while improving both accuracy and computational efficiency.
The previous FastConformer–Transformer stack is streamlined into a pure Transformer encoder.
A learned upsampling layer then maps the encoder’s coarse 80-ms representations to 10-ms resolution, enabling fine-grained predictions.
Training leverages large-scale simulated mixtures and adds activity and phantom-speaker losses to the standard Sortformer objectives.
Despite doubling speaker capacity over its predecessor, the new architecture reduces training memory by a factor of 2.8, makes training 2.2× faster, and delivers up to 6.4× higher inference throughput.
Nemotron 3 Diarization establishes a new state of the art in DER on DIHARD III and CALLHOME.
When used to condition DiCoW, it also achieves state-of-the-art tcpWER on the single-microphone conditions of AMI and NOTSOFAR-1.
Training code is open source, and model weights are released under the permissive OpenMDW-1.1 license.
\end{abstract}


\section{Introduction}
\label{sec:introduction}

Speaker diarization remains a core capability for meeting transcription and multi-talker automatic speech recognition (ASR), even as speech large language models~(LLMs) increasingly generate speaker-attributed transcripts directly. Recent systems and benchmarks explore unified prediction of lexical content, timestamps, and speaker labels \cite{mu2026mlcslm,peng2026vibevoice}. Yet these developments do not eliminate the need for explicit diarization: DM-ASR~\cite{li2026dmasr} uses diarization as an explicit structural prior for an LLM, while a recent comparative study~\cite{tawara2026whospoke} found unified models competitive in two-speaker conditions but less robust than modular pipelines as speaker count and overlap increase. Accurate diarization therefore remains valuable both as a standalone capability and as a component of systems built around speech LLMs.

End-to-end neural diarization (EEND) replaces multi-stage diarization pipelines with direct prediction of joint speaker activity and naturally represents overlapping speech \cite{fujita2019eend}. Extensions based on encoder-decoder attractors accommodate an unknown number of speakers~\cite{horiguchi2020eda}, while EEND with vector clustering combines overlap-aware local predictions with global embedding clustering to handle long recordings and arbitrary speaker inventories \cite{kinoshita2021eendvc}. This local-to-global design underlies widely used systems such as Pyannote~\cite{bredin2023pyannote} and DiariZen~\cite{han2025diarizen,han2025efficient}. By contrast, Sortformer~\cite{park2025sortformer} follows a fixed-capacity EEND approach, combining permutation-invariant loss~(PIL) with arrival-time-sort~(ATS) supervision to assign speakers to output channels according to their order of first arrival. Because densely annotated real conversations are scarce, EEND relies heavily on simulated mixtures. Large-scale pre-training improves performance~\cite{broughton2025limits}, but its effectiveness depends on turn-taking statistics, source diversity, acoustic augmentation, and how simulated and real data are combined~\cite{polok2026fastmss}.

Most of this progress targets offline processing. Importantly, streaming diarization must preserve identities across chunks despite bounded right context, overlapping speech, and newly arriving speakers. Prior online systems address this problem through limited-context encoders, speaker-tracing buffers, or blockwise speaker detection and representation \cite{han2021bweend,xue2021flexstb,horiguchi2022online,cheng2025s2snd}. Streaming Sortformer~\cite{medennikov2025streaming} adds an arrival-order speaker cache (AOSC) to maintain channel ordering across chunks, allowing one model to serve multiple latency profiles. Its released model, however, is limited to four speakers and produces predictions only every 80~ms. This leaves a gap for an open model that combines competitive accuracy in both offline and low-latency streaming settings with support for more speakers, precise boundaries, and high computational efficiency.

These requirements are tightly coupled. Within the AOSC design, supporting more speakers requires a larger cache to retain sufficient evidence for each identity, which directly increases inference cost. Additional output channels also remain unused in typical recordings, increasing the risk of phantom-speaker activations. Meanwhile, running the encoder at a 10-ms frame step to improve boundary precision would greatly increase sequence length and attention cost. The model must also remain robust across latency profiles that provide different amounts of future context. These trade-offs motivate a design that supports more speakers and finer temporal resolution without sacrificing streaming efficiency.

This work introduces \model\footnote{\scriptsize{\url{https://hf.co/nvidia/Nemotron-3-Diarization}}}, an open-weight model addressing these constraints. Our contributions are threefold. First, we redesign the architecture to efficiently increase speaker capacity and output temporal resolution.
Second, we introduce an \emph{activity loss} that models overall speech activity and a \emph{phantom loss} that suppresses false activations in unused speaker channels. We pair these objectives with mixed-data and variable-context training to support both offline and streaming operation.
Third, we analyze these choices through ablations of the backbone, initialization, temporal resolution, objectives, augmentations, and training data. We evaluate diarization accuracy across latencies and benchmark training and inference efficiency. We also assess downstream speaker-attributed ASR in offline and streaming settings.

\clearpage
\section{Nemotron 3 Diarization}
\label{sec:model}

\subsection{Training data}

\model~is trained on approximately 5,000~h each of human-annotated and pseudo-labeled real recordings, together with around 82,600~h of simulated multi-talker mixtures.

The real multi-domain training data include the following public corpora: AISHELL-4~\cite{fu2021aishell4}, AliMeeting~\cite{yu2022m2met}, AMI~\cite{carletta2005ami}, CALLHOME Part~1~\cite{canavan1997callhome}, DIHARD III Development~\cite{ryant2021dihard}, DiPCo~\cite{vansegbroeck2019dipco}, DISPLACE-2024~\cite{kalluri2024displace}, DISPLACE-M~\cite{dhanya2026displacem}, Fisher English~\cite{cieri2004fisher}, ICSI~\cite{janin2003icsi}, NOTSOFAR-1~\cite{vinnikov2024nots} (NSF1), and VoxConverse-v0.3~\cite{chung2020voxconverse}.
Additionally, a licensed David~AI\footnote{\scriptsize{\url{https://www.withdavid.ai/\#datasets}}} dataset of around 1,000 hours of 3--4-speaker English conversations is used.
We also include a 5,000-hour multi-talker subset of YODAS-v2~\cite{li2023yodas} pseudo-labeled using an ensemble of nine Sortformers combined via DOVER-Lap~\cite{raj2021doverlap}.

The simulated mixtures (1--8 speakers, 180--200~s per mixture) were created using the FastMSS\footnote{\scriptsize{\url{https://github.com/popcornell/FastMSS}}} toolkit following the \emph{Mind the Gap} protocol from~\cite{polok2026fastmss}. 
The simulated data comprise 6,694~h from LibriSpeech~\cite{panayotov2015librispeech}, 6,743~h from AMI headsets, 6,745~h from AliMeeting headsets, 6,755~h from Fisher English, 36,458~h from around 22,000~h of licensed David AI English audio, and 19,216~h from around 2,000~h of licensed David AI multilingual audio in 21 languages. 
Notably, simulations use tight speaker-activity regions obtained with the Montreal Forced Aligner~\cite{mcauliffe2017mfa}.

\subsection{Architecture}
\model~builds on the Streaming Sortformer~\cite{medennikov2025streaming} paradigm with several important changes.
First, the backbone is streamlined from a 17-layer FastConformer~\cite{rekesh2023fastconformer} followed by an 18-layer Transformer~\cite{vaswani2017attention} to a single 31-layer Transformer with rotary positional embeddings~(RoPE)~\cite{su2024roformer}.
This change is motivated by both higher computational efficiency and better scalability of a pure Transformer. This also makes the model more compact, reducing the total number of parameters from 117M to 99M. 
Input features are downsampled eightfold through feature stacking. The Transformer encoder therefore retains the FastConformer’s 80-ms frame step.

Second, we introduce a learned one-dimensional subpixel convolution on top of the encoder. This layer upsamples the encoder outputs by a factor of eight, enabling accurate detection of short turns and precise boundaries while preserving the efficient 80-ms encoder frame step.
Specifically, a width-3 convolution maps each \(H\)-dimensional encoder state to \(8H\) channels, which are rearranged into eight consecutive \(H\)-dimensional states. The encoder therefore remains at 80-ms resolution, while the output heads operate at an effective 10-ms resolution with almost no computational overhead.

Finally, AOSC capacity is increased from 188 to 264 frames to accommodate the twofold increase in supported speaker count. The mean silence embedding used to separate speakers in AOSC is replaced with a learnable embedding for greater robustness.

\subsection{Training objectives}

Alongside the standard Sortformer PIL and ATS losses, which use the best target-speaker permutation and arrival-time-ordered targets, respectively, \model~introduces two auxiliary losses: the \emph{activity loss} and the  \emph{phantom loss}.

The \emph{activity loss} is a cross-entropy loss over three permutation-free frame activity classes: silence, single-speaker speech, and overlapping speech. A separate auxiliary head predicts these classes from the encoder representations. 
This auxiliary task regularizes training by supervising overall speech activity without requiring speaker assignment.

The \emph{phantom loss} addresses speaker overcounting caused by high-confidence
activity predictions (``phantom speakers'') in channels empty under both the PIL
and ATS alignments.
Let \(p_{bts}\) be the speech posterior and \(\operatorname{sg}\) denote stop-gradient.
For each empty channel, \(\mathcal{I}_{bs}\) contains valid frames satisfying
\(\operatorname{sg}(p_{bts})>\theta\); otherwise,
\(\mathcal{I}_{bs}=\varnothing\).
Each selected frame incurs the inactive-label negative log-likelihood
\(\ell_{bts}=-\log(1-p_{bts})\).
These losses are pooled as
\begin{equation*}
\mathcal{L}_{\mathrm{ph}}
=\frac{\tau}{BS}\sum_{b=1}^{B}\sum_{s=1}^{S}
\log\!\left[
\frac{1}{|\mathcal{I}_{bs}|}
\sum_{t\in\mathcal{I}_{bs}}
\exp\!\left(\frac{\ell_{bts}}{\tau}\right)
\right].
\end{equation*}
We set \(\theta=0.25\), \(\tau=0.5\) and \(S=8\), and define the contribution as zero when \(\mathcal{I}_{bs}=\varnothing\).
This normalized log-mean-exp pooling emphasizes the largest false activations without scaling directly with the number of selected frames.

\subsection{Model training and augmentations}

Training is performed using the NVIDIA NeMo Speech Framework\footnote{\scriptsize{\url{https://github.com/NVIDIA-NeMo/Speech}}}~\cite{kuchaiev2019nemo} on 64 NVIDIA A100-SXM4-80GB GPUs in two stages.

First, offline Sortformer training~\cite{park2025sortformer} is performed on simulated mixtures only, with the encoder initialized from a self-supervised NEST~\cite{huang2025nest} Transformer pretrained on a 900,000-hour subset of Granary-v2, an expanded version of the Granary corpus~\cite{koluguri2025granary}. 
Training uses random 60-second chunks drawn on the fly from full recordings.
The PIL, ATS, \emph{activity}, and \emph{phantom} loss weights are 0.5, 0.5, 0.5, and 0, respectively.
The offline stage operates at the native 80-ms output resolution without the upsampling layer.

Second, the model is fine-tuned for streaming with AOSC~\cite{medennikov2025streaming} enabled, using both real and simulated data.
The 105-second training samples are constructed on the fly, with two chunks from different parts of a source recording concatenated with probability 0.5; sampling favors examples containing more active speakers.
The PIL, ATS, \emph{activity}, and \emph{phantom} loss weights are 0.75, 0.25, 0.5, and 0.1, respectively. The upsampling layer is introduced at this stage, initialized to repeat each encoder state across eight output positions, and jointly trained with the rest of the model using 10-ms targets.

Additionally, to prevent the model from relying on future context during streaming inference, each batch receives a \emph{block-causal tail} with probability 0.5. This modifies the self-attention mask to contain a bidirectional prefix followed by a randomly sampled tail. The tail is divided into blocks of one to eight frames: attention is bidirectional within each block, while each block may attend to the prefix and earlier blocks but not to future blocks. Full-context batches and this family of masks jointly expose one checkpoint to offline and multiple streaming contexts.

Finally, both training stages apply SpecAugment~\cite{park2019specaugment} regularization with two frequency masks spanning up to 24 feature bins each. 
The detailed training parameters used are given in the publicly available offline\footnote{\scriptsize{\url{https://github.com/NVIDIA-NeMo/Speech/blob/main/examples/speaker_tasks/diarization/conf/neural_diarizer/sortformer_offline_8spk.yaml}}} and streaming\footnote{\scriptsize\url{https://github.com/NVIDIA-NeMo/Speech/blob/main/examples/speaker_tasks/diarization/conf/neural_diarizer/sortformer_streaming_8spk.yaml}} configuration files in the NeMo Speech repository.

\begin{table*}[!th]
\caption{DER (\%) $\downarrow$ in offline and streaming conditions across evaluation sets. DH3, CH, Ali and NSF1 denote DIHARD~III Eval, CALLHOME Part~2, AliMeeting and NOTSOFAR-1, respectively. All overlap is scored; collar tolerance is 0.25~s for CH and 0~s for the other sets. Best offline results are bold.}
\label{tab:der}
\centering
\small
\setlength{\tabcolsep}{3.2pt}
\begin{tabular}{lccccccccc}
\toprule
System & Latency & DH3 & CH & Ali-Near & Ali-Far & AMI-MHM & AMI-SDM & NSF1-MHM & NSF1-SC \\
\midrule
DiariZen & offline
  & 14.5 & 14.68 & 16.78 & 10.8$^{*}$ & 15.56 & 13.9$^{*}$ & 14.20 & 16.7$^{*}$ \\
Sortformer-v2.1 & 30.4~s
  & 19.09 & 10.32 & 11.57 & 13.69 & 15.81 & 21.42 & 21.77 & 30.49 \\
\model & 30.4~s
  & \textbf{12.73} & \textbf{9.10} & \textbf{6.40} & \textbf{10.47} & \textbf{9.25} & \textbf{11.14} & \textbf{6.77} & \textbf{11.00} \\
\midrule
Sortformer-v2.1 & 1.04~s
  & 19.60 & 11.31 & 12.47 & 15.58 & 16.36 & 21.73 & 22.12 & 31.81 \\
\model & 1.04~s
  & 13.18 & 10.29 & 6.59 & 10.80 & 9.48 & 12.80 & 7.70 & 12.77 \\
\midrule
Sortformer-v2.1 & 0.32~s
  & 19.85 & 12.67 & 13.68 & 16.85 & 17.77 & 23.89 & 23.37 & 32.95 \\
\model & 0.32~s
  & 13.55 & 11.32 & 7.19 & 11.60 & 10.05 & 12.95 & 8.65 & 14.53 \\
\midrule
SOTA 09/2026 & offline
  & 14.1~\cite{palka2026spherevbx} & 10.08~\cite{chen2024attention} & 8.60$^{*}$~\cite{broughton2025limits} & 10.6$^{*}$~\cite{palka2026spherevbx} & 11.06$^{*}$~\cite{broughton2025limits} & 13.7$^{*}$~\cite{palka2026spherevbx} & --- & 16.5$^{*}$~\cite{palka2026spherevbx} \\
\bottomrule
\end{tabular}

\vspace{1mm}
\parbox{0.98\textwidth}{\scriptsize $^{*}$Reported using the original/default reference annotations rather than the force-aligned reference annotations used for the corresponding results in this work; these values are therefore not directly comparable with ours.}
\end{table*}

\section{Results}
\label{sec:results}

\subsection{Diarization accuracy evaluation}
\label{ssec:accuracy}

The evaluation contains the following datasets: DIHARD~III evaluation~\cite{ryant2021dihard} (259 recordings, 1--9 speakers per recording), CALLHOME Part~2 (250 recordings, 2--6 speakers), AliMeeting~\cite{yu2022m2met} near and far microphones (20 recordings, 2--4 speakers), AMI~\cite{carletta2005ami} Mix of Headset Microphones~(MHM) and Single Distant Microphone~(SDM) conditions (16 recordings, 3--4 speakers), and NOTSOFAR-1~\cite{vinnikov2024nots} MHM and Single-Channel~(SC) conditions (160 recordings, 3--7 speakers). All overlapped speech is scored. DER uses no forgiveness collar except for CALLHOME, where the conventional 0.25-s collar is used. 
It is important to note that we use forced-alignment-based reference labels for AMI, AliMeeting\footnote{\scriptsize{\url{https://github.com/nttcslab-sp/diar-forced-alignment}}} and NOTSOFAR-1\footnote{\scriptsize{\url{https://github.com/popcornell/FastMSS/blob/master/resources/notsofar1-channels_mfa_rttms.tar.gz}}} instead of the original segment-level annotations that label substantial within-segment silences as speech, thereby overestimating reference speaker activity~\cite{horiguchi2025can}. 

We consider two open-source baselines: the previous four-speaker Streaming Sortformer-v2.1\footnote{\scriptsize\url{https://hf.co/nvidia/diar_streaming_sortformer_4spk-v2.1}}~\cite{medennikov2025streaming} and DiariZen\footnote{\scriptsize\url{https://hf.co/BUT-FIT/diarizen-wavlm-large-s80-md-v2}}~\cite{han2025diarizen,han2025efficient}, the strongest offline system.
Table~\ref{tab:der} reports DER at the three input-buffer latencies of 30.4, 1.04, and 0.32~s used in the original Streaming Sortformer study~\cite{medennikov2025streaming}. Reducing the buffer from 30.4 to 0.32~s increases DER by 0.79--3.53 percentage points across the evaluation sets, thus demonstrating graceful degradation. The same checkpoint can operate with input buffers as short as 0.08~s, although accuracy degrades further; we omit this regime to retain direct comparability with the previous work. At all three reported latencies, \model~outperforms Sortformer-v2.1 across every evaluation set and the offline DiariZen baseline on every directly comparable condition.
This establishes a new state of the art~(SOTA) on DIHARD III and CALLHOME benchmarks. 
Although we believe our results on AMI, AliMeeting, and NOTSOFAR-1 are also SOTA, we do not claim this formally because the starred results use different ground-truth labels and are therefore not directly comparable with ours.

\subsection{Training and inference efficiency}

During training on 64 NVIDIA A100 GPUs in \texttt{bf16-mixed} precision, the streamlined Transformer reduces memory consumption by a factor of 2.8 and iteration time by a factor of 2.2 relative to an otherwise identical model with the legacy FastConformer encoder.

\begin{table}[t]
\caption{Inference RTFx $\uparrow$ on an RTX PRO 5000 Blackwell GPU using \texttt{bf16}; ``comp.'' denotes dynamic \texttt{torch.compile}. The best result for each batch size and latency is bold.}
\label{tab:rtfx}
\centering
\setlength{\tabcolsep}{2.0pt}
\begin{tabular}{@{}llrrrr@{}}
\toprule
& & \multicolumn{2}{c}{Batch 1} & \multicolumn{2}{c}{Batch 32} \\
\cmidrule(lr){3-4}\cmidrule(lr){5-6}
Model & Latency & eager & comp. & eager & comp. \\
\midrule
DiariZen        & offline & 20 & -- & 44 & -- \\
Sortformer-v2.1 & 30.4~s & 874 & 1468 & 3204 & 2619 \\
\model & 30.4~s & 1340 & \textbf{4385} & 12196 & \textbf{15113} \\
\midrule
Sortformer-v2.1 & 1.04~s & 16 & 42 & 193 & 136 \\
\model & 1.04~s & 38 & \textbf{164} & 581 & \textbf{865} \\
\midrule
Sortformer-v2.1 & 0.32~s & 8 & 21 & 101 & 76 \\
\model & 0.32~s & 12.5 & \textbf{54} & 199 & \textbf{292} \\
\bottomrule
\end{tabular}
\end{table}

Table~\ref{tab:rtfx} compares inference speed on one NVIDIA RTX PRO 5000 Blackwell GPU in \texttt{bf16} precision. Here, Real-Time Factor Speedup~(RTFx) defined as audio duration divided by wall-clock processing time is calculated with batch sizes of 1 and 32 on a large set of around 900~s long recordings. Compiled measurements exclude the one-time graph compilation overhead and are collected after compilation and warm-up.
It reveals massive inference speed improvements over Sortformer-v2.1 with legacy FastConformer~\cite{rekesh2023fastconformer} encoder, despite the twofold increase in speaker capacity. Notably, even with the ultra-low input buffer latency of 0.32~s, the model is fast enough to support multiple parallel sessions on a single GPU.

\subsection{Ablations}

\begin{table}[t]
\caption{Ablations on the compound evaluation set: globally pooled, zero-collar DER~(\%) $\downarrow$ at input-buffer latencies of 30.4~s and 0.32~s. $\Delta$ is the absolute DER change; positive values indicate degradation.}
\label{tab:ablation}
\centering
\setlength{\tabcolsep}{4.0pt}
\begin{tabular}{lcccc}
\toprule
& \multicolumn{2}{c}{30.4~s} & \multicolumn{2}{c}{0.32~s} \\
\cmidrule(lr){2-3}\cmidrule(lr){4-5}
System & DER & $\Delta$ & DER & $\Delta$ \\
\midrule
\model                               & 10.42 & --    & 12.06 & -- \\
\quad FastConformer enc.             & 11.36 & +0.94 & 13.33 & +1.27 \\
\quad no NEST initialization         & 12.05 & +1.63 & 14.28 & +2.22 \\
\quad downsampled (80~ms)            & 11.09 & +0.67 & 12.67 & +0.61 \\
\quad no upsampling (80~ms)          & 12.69 & +2.27 & 14.09 & +2.03 \\
\quad no phantom loss                & 10.41 & -0.01 & 12.15 & +0.09 \\
\quad no activity loss               & 11.42 & +1.00 & 13.03 & +0.97 \\
\quad no SpecAugment                 & 10.88 & +0.46 & 12.47 & +0.41 \\
\quad no block-causal tail           & 10.50 & +0.08 & 12.36 & +0.30 \\
\quad no simulated data              & 13.40 & +2.98 & 15.21 & +3.15 \\
\quad no David AI data               & 11.19 & +0.77 & 12.83 & +0.77 \\
\quad no YODAS data                  & 10.81 & +0.39 & 12.38 & +0.32 \\
\bottomrule
\end{tabular}
\end{table}

To assess the contribution of each component, Table~\ref{tab:ablation} reports DER and its absolute change relative to \model~at offline~(30.4~s) and ultra-low~(0.32~s) input buffer latency. All recordings from the constituent evaluation sets are scored jointly as one compound evaluation set with a zero collar.
Each ablation is represented by a single training run using the same optimization and evaluation settings. We therefore report observed effect sizes rather than statistical significance and interpret small differences cautiously. We emphasize effects that are consistent across both latency profiles or appear in the metric directly targeted by the component.

The pure Transformer backbone outperforms the legacy FastConformer. This may reflect better scaling with the available data or better compatibility with AOSC: cached speaker blocks do not form a continuous temporal sequence, so local convolutions can mix unrelated states across artificial block boundaries. The substantial degradation without NEST initialization highlights the importance of broad self-supervised pretraining, suggesting that the diarization data are insufficient to learn equally robust acoustic representations from scratch.

Removing high-resolution upsampling degrades DER considerably more than downsampling the 10-ms predictions to 80~ms before scoring. This indicates that learning with high-resolution outputs provides benefits beyond finer evaluation granularity.

Removing the phantom loss leaves DER nearly unchanged but increases speaker-count MAE from 0.157 to 0.222 offline and from 0.270 to 0.491 at 0.32-s latency, confirming its role in reducing overcounting. Removing the activity loss or SpecAugment also degrades DER at both latency profiles. The block-causal-tail augmentation has little effect at 30.4~s but provides a clearer benefit at 0.32~s.

Finally, training without simulated data produces the largest degradation, underscoring the importance of simulated mixtures. David AI and pseudo-labeled YODAS data also contribute, as excluding either degrades DER at both latency profiles.

\subsection{Downstream multi-talker ASR performance}

As already highlighted in subsection~\ref{ssec:accuracy}, DER is a highly annotation-style-dependent metric, which makes it hard to compare diarization systems trained on different labels directly. To overcome this limitation, we conduct a comparison of our \model~with the baseline systems in a downstream multi-talker automatic speech recognition~(MT-ASR) task. Unlike DER, time-constrained minimum-permutation Word Error Rate~(tcpWER) operates on speaker-attributed ASR hypotheses rather than the diarization labels used to score frame-level errors. We calculate tcpWER using the \texttt{meeteval} package~\cite{vonneumann2023meeteval} with the default collar tolerance of 5~s and the \emph{chime8} text normalizer.

Table~\ref{tab:mtasr} therefore evaluates whether the diarization improvements transfer across reference-annotation protocols and two different MT-ASR models: a state-of-the-art offline DiCoW-v3.3-large\footnote{\scriptsize{\url{https://hf.co/BUT-FIT/DiCoW_v3_3_large}}}~\cite{polok2026dicow,polok2026sedicow}, and a streaming Multitalker-Parakeet-v1\footnote{\scriptsize{\url{https://hf.co/nvidia/multitalker-parakeet-streaming-0.6b-v1}}}~\cite{wang25y_interspeech} operating at default 1.12~s input buffer latency.

\begin{table}[t]
\caption{Downstream tcpWER (\%) $\downarrow$ for streaming and offline Multi-Talker ASR. Hard and soft denote segment-level and posterior-based diarization conditioning, respectively. Bold indicates the best non-oracle result.}
\label{tab:mtasr}
\centering
\setlength{\tabcolsep}{3.0pt}
\begin{tabular}{@{}lcccc@{}}
\toprule
& \multicolumn{2}{c}{AMI} & \multicolumn{2}{c}{NSF1} \\
\cmidrule(lr){2-3}\cmidrule(lr){4-5}
Diarization & MHM & SDM & MHM & SC \\
\midrule
\multicolumn{5}{@{}l}{\textit{Streaming: Multitalker-Parakeet-v1 (1.12~s latency)}} \\
Ground truth (oracle) & 12.66 & 17.79 & 16.86 & 26.34 \\
Sortformer-v2.1 & 18.64 & 34.13 & 35.59 & 51.35 \\
\model~(hard) & \textbf{13.84} & \textbf{21.18} & \textbf{18.68} & \textbf{31.65} \\
\midrule
\multicolumn{5}{@{}l}{\textit{Offline: DiCoW-v3.3-large (greedy decoding)}} \\
Ground truth (oracle) & 11.72 & 15.27 & 8.78 & 15.45 \\
DiariZen & 15.17 & 19.32 & 18.57 & 25.63 \\
Sortformer-v2.1 & 18.17 & 30.11 & 30.14 & 42.56 \\
\model~(hard) & 12.65 & 17.19 & 10.47 & 19.23 \\
\model~(soft) & \textbf{12.35} & \textbf{17.06} & \textbf{10.30} & \textbf{18.95} \\
\midrule
SOTA 09/2026 & 15.2~\cite{polok2026sedicow} & 18.5~\cite{polok2026sedicow} & N/A & 23.63~\cite{niu2024ustc} \\
\bottomrule
\end{tabular}
\end{table}

DiCoW conditioned on segment-level~(hard) predictions from \model~substantially outperforms both baselines, while posterior-based~(soft) conditioning further reduces tcpWER in all four conditions. To our knowledge, this is the first consistent soft-over-hard gain with DiCoW. The resulting system also establishes new state-of-the-art tcpWER on both AMI and NOTSOFAR-1.
In particular, on the NOTSOFAR-1 single-channel condition, the resulting system surpasses USTC-NERCSLIP \textit{sys2}~\cite{niu2024ustc}~--- the winner of the CHiME-8 NOTSOFAR-1 Challenge~\cite{vinnikov2024nots}\footnote{The 22.2\% quoted in the paper and leaderboard is a macro-average, not the compound (globally pooled) tcpWER used in our comparison.}.

With streaming Multitalker Parakeet, \model~also substantially reduces tcpWER relative to the Sortformer-v2.1 baseline. The consistent reductions in tcpWER across both ASR back-ends show that the diarization improvements transfer to downstream transcription rather than merely reflecting better agreement with the forced-aligned reference labels used for DER.

\section{Conclusion}
\label{sec:conclusion}

\model~extends the Streaming Sortformer framework from four to eight speakers while significantly improving both accuracy and throughput. Notably, it establishes a new state of the art on DIHARD III and CALLHOME in terms of DER, and on AMI and NOTSOFAR-1 in terms of tcpWER. Given that the model is released under the permissive OpenMDW-1.1 license and is easily fine-tunable on target-domain data, we hope it sees wide adoption by both researchers and enterprises. 
Future work will further extend the supported speaker count, scale the model using expanded synthetic and pseudo-labeled corpora, and investigate direct integration into multimodal speech LLMs to enable seamless, end-to-end conversational reasoning.

\clearpage
\bibliographystyle{IEEEbib}
\bibliography{references}

\end{document}